\documentclass{article}
\usepackage{arxiv}

\usepackage[T1]{fontenc}	
\usepackage[utf8]{inputenc}
\usepackage[british]{babel}
\usepackage{amsmath,amssymb}
\usepackage{setspace}
\usepackage{booktabs}
\usepackage{xcolor}
\usepackage{microtype}
\usepackage[hyphens]{url}
\usepackage{hyperref}

\providecommand{\tightlist}{%
  \setlength{\itemsep}{0pt}\setlength{\parskip}{0pt}}

\hypersetup{
  pdftitle={After Cheap Discovery: From Unknown to Known-and-Unfixed},
  pdfauthor={Bahman Sistany},
  pdflang={en-GB},
  pdfkeywords={vulnerability management, security economics, remediation
    coverage, software liability, formal verification},
  colorlinks=true,
  linkcolor={black},
  citecolor={Blue},
  urlcolor={black}}

\title{After Cheap Discovery: From Unknown to Known-and-Unfixed}

\usepackage{etoolbox}
\makeatletter
\providecommand{\subtitle}[1]{%
  \apptocmd{\@title}{\par {\large #1 \par}}{}{}
}
\makeatother
\subtitle{Automated discovery has collapsed the cost of finding
vulnerabilities. The binding constraint has moved to what happens after
a finding is recorded.}

\renewcommand{\shorttitle}{After Cheap Discovery}
\renewcommand{\headeright}{Preprint}
\renewcommand{\undertitle}{Preprint}

\author{%
  Bahman Sistany \\
  Independent Researcher \\
  Ottawa, Canada \\
  \texttt{bahman@sistany.com} \\
}

\date{2 September 2026}

\begin{document}
\maketitle
\begin{abstract}
Automated vulnerability discovery has removed the scarcity of expert
attention that protected most software. The response has concentrated on
discovery and on repair, and both are becoming cheaper. This article
argues that neither cost curve determines exposure. What determines it
is remediation coverage at the release decision: the fraction of
identified vulnerabilities fixed before a product ships, and the residue
of known, assessed, unremediated flaws an organisation has decided to
ship with. Four arguments follow. The residue is not a random sample of
what was found, because triage sorts on cost and the expensive cases are
architectural. The deferred backlog is itself a high-value artifact.
Documented awareness alters an organisation's legal and market position,
and produces an adverse selection that the price of software does not
reflect. And
no regulatory instrument reaching vendors triggers on internal knowledge
--- every one fires on exploitation observed by a third party --- which
leaves the distance between what a vendor knows and what it must
disclose entirely at the vendor's discretion. CISA's Binding Operational
Directive 26-04 is examined as the exception that shows what a solution
requires: a regulator's reach over a decision is bounded by its
visibility of the input to that decision, so a duty attaching to
internal knowledge needs an independent record of that knowledge to
attach to. Release coverage is not currently measured, and measuring it
is the precondition for liability, insurance or procurement to act.
\end{abstract}


\setstretch{1.12}
Two results this year are worth taking as settled before anything else
is argued.

In April, Anthropic announced Claude Mythos Preview alongside Project
Glasswing, a limited-access consortium formed because broad release was
judged unwise. The company's red team reported {[}1{]} that the model
could find and then exploit zero-day vulnerabilities in every major
operating system and browser when directed to do so. The UK's AI
Security Institute evaluated {[}2{]} the model independently. It
observed multi-stage attacks and autonomous discovery under controlled
conditions. Two limits are worth recording: the target networks were
poorly defended, and the model paid no penalty for noisy behaviour.

In July, OpenAI disclosed {[}3{]} an incident from an internal
evaluation of two models. The evaluation was run without the production
classifiers that normally stop a model pursuing high-risk cyber
activity, so that maximal capability could be measured. Both left their
sandbox by exploiting a previously unknown flaw in permitted egress
infrastructure. From there they chained further vulnerabilities to reach
Hugging Face's production database. Neither model had been directed to
attack anything. Both were pursuing a benchmark score.

These results are limited in the ways their authors say they are, and
those limits do not touch what matters here. Taken together the two
results establish one thing: finding exploitable flaws in real software
no longer requires scarce expertise, and no longer requires that a human
choose the target.

One further limit does need conceding. Both systems were gated ---
Mythos Preview behind a consortium, the OpenAI models internal --- so
the results establish that the capability exists, not the price at which
it is generally available. The argument that follows does not turn on
that price. It turns on defenders finding more than they can fix, and
commodity scanning already delivers that without any of this.

This article concerns what follows from that, and only that. It is not
about whether the capability generalises, nor about what attackers will
do with it. Discovery produces a finding; the finding is triaged; some
fraction is fixed before the product ships. This article is about the
third step, which has received much less attention and which I will
argue is now the constraint.

\hypertarget{an-older-question-and-what-has-changed-about-it}{%
\section{An older question, and what has changed about
it}\label{an-older-question-and-what-has-changed-about-it}}

The question of what vulnerability research buys us is not new. Ross
Anderson argued in 2001 {[}4{]} that information insecurity is at least
as much a matter of perverse incentives as of technical deficiency, and
that the useful vocabulary is microeconomic: externalities, asymmetric
information, moral hazard, adverse selection. Three years later Eric
Rescorla asked at the Workshop on the Economics of Information Security
whether finding security holes is a good idea at all {[}5{]}, observing
that the case for the activity depends on assumptions that may not
survive examination. If bugs are dense, removing one leaves the
population essentially unchanged. If the bug you find would have been
found by someone else anyway, the finding buys less than it appears to.

Adam Chlipala has recently given the sharpest statement {[}6{]} of what
that expert effort was actually protecting. His argument is that the
protection most software enjoyed was never technical but allocational: a
small population of experts spent their hours where the zero-day market
made the spending worthwhile, leaving the majority of code defended by
nothing more than not being worth the trouble. That amounted to a
subsidy paid by the scarcity of expert hours. Automating discovery
dissolves the constraint and withdraws the subsidy with it. He then
argues that the same collapse in cost lets defenders scan their own code
before shipping it. Beyond that he makes the case for formal
verification, on the grounds that a proof closes a category of bug
permanently rather than one instance at a time.

I take that account to be correct, and this article assumes it rather
than arguing with it. What I want to examine is a step it does not
reach.

\hypertarget{defining-the-quantity-of-interest}{%
\section{Defining the quantity of
interest}\label{defining-the-quantity-of-interest}}

Before, an unfixed flaw was usually one nobody had found. Some had been
found and deferred, and those looked much as they do now, but they were
few enough that an organisation could treat them as a backlog. Now
someone looks,
records it, assigns a severity, and a release proceeds anyway. The
technical exposure is identical in both cases. Its epistemic status is
not, and neither is its legal status. The field already has a name for
the quantity that measures the difference.

Remediation coverage is the proportion of detected vulnerabilities that
have been successfully remediated, and it can be scoped to a severity
band where that is useful:

\begin{equation}\label{eq:coverage}
\text{coverage} \;=\;
\frac{\text{vulnerabilities remediated}}{\text{vulnerabilities identified}}
\end{equation}

Both counts are already held by most organisations. The ratio is rarely
computed at the point where it carries the most information, which is
the release decision. The quantity this article is concerned with is the
difference between the two counts:

\begin{equation}\label{eq:residue}
\text{residue} \;=\;
\text{vulnerabilities identified} \;-\; \text{vulnerabilities remediated}
\end{equation}

That difference is not an error term. It is the set of vulnerabilities
the organisation knows about, has assessed, and has decided to ship
with.

Coverage is the rate at which findings convert. The residue is the stock
that accumulates when they do not. The rest of this article uses both,
because the arguments about incentives concern the rate and the
arguments about exposure concern the stock.

Two counting rules have to be stated, because a ratio proposed as a
procurement artifact cannot leave its own terms open. A finding is
\emph{identified} when it has been confirmed, deduplicated, and
attributed to the release under decision. Raw scanner output is not the
denominator: an unfiltered feed measures the tool's false-positive rate
rather than the organisation's exposure, and an organisation running
five overlapping scanners would post a worse figure than one running a
single tool well. The confirmation step is where a finding acquires the
epistemic status this article is about, so it is also the right place to
start counting. A finding is \emph{remediated} when the flaw is no
longer present in the shipped artifact. Mitigations do not qualify. A
finding that has been mitigated but not fixed counts in the denominator
of Equation~\ref{eq:coverage} and not in the numerator, and belongs in
the residue of Equation~\ref{eq:residue}, because a mitigation is a
decision to ship with the flaw present; where mitigation is material it
should be reported as a separate named line rather than folded into
either count. Both rules are conventions and other defensible ones
exist, but a coverage figure computed under an undeclared convention is
not comparable to any other, which defeats the purpose of computing it.

Stating the denominator invites the obvious attack on it. Coverage is a
ratio, so it improves when the numerator rises and equally when the
denominator falls, and the denominator falls whenever an organisation
narrows a scan scope, retires a tool, or raises a confidence threshold.
This is Rescorla's objection in miniature, arriving a step earlier than
he raised it: any measure that rewards conversion penalises looking. The
defence available is not a better formula but a disclosure requirement.
A coverage figure is uninterpretable on its own and should be reported
as a pair: the ratio, and the scan configuration and scope that produced
it. Under that convention a decline caused by better instrumentation is
explicable, and an improvement caused by reduced instrumentation is
visible as a change in the declared configuration rather than as
progress. This puts weight on the configuration record, which is
consequently the thing an auditor or a buyer should examine first. The
weakness remains: coverage
reported without a declared scope can be manufactured, and nothing in
the arithmetic prevents it.

It is worth distinguishing coverage from the metrics it is usually
reported alongside, because they answer different questions. Mean time
to remediate measures elapsed time for the items that were fixed. Escape
rate measures the share of vulnerabilities first discovered in
production rather than earlier in the pipeline. Fix rate comes closest:
findings closed against findings opened in a period is coverage by
another name, and it is widely tracked. But it is an estate-level figure
computed over a window, and none of these reports the fraction of known
findings that were never addressed at all before a particular product
shipped.

The word coverage will do three jobs in what follows, and it is worth
marking them now. Estate coverage is the rolling operational figure.
Release coverage is the fraction converted before a particular product
ships, and it is this article's subject. Deployment coverage is the
fraction of published vulnerabilities an operator has patched in what it
runs, and it is where the only usable public data happens to be.

Release coverage is narrower than estate coverage, and the difference
matters: an operational dashboard
treats an unremediated finding as work outstanding, whereas a release
gate treats it as a decision already taken. The measure presupposes a
release boundary, which versioned product supplies and continuously
deployed software does not; where there is no ship date the boundary has
to be drawn at whatever gate a human actually signs, which is usually a
change-approval step or a promotion to production, and the argument
below holds wherever such a gate exists rather than only where a version
number does.

Both responses to cheap discovery terminate at this quantity.
Pre-release scanning produces findings, which must convert. Verification
produces proof obligations, which must either be discharged or the
specification weakened. In both cases the final step is a decision made
under a deadline about whether to act on something now known.
The question of when mitigation is an adequate substitute for
remediation is outside this article's scope.

\hypertarget{whether-repair-is-also-getting-cheaper}{%
\section{Whether repair is also getting
cheaper}\label{whether-repair-is-also-getting-cheaper}}

The argument so far assumes that fixing remains expensive relative to
finding. That assumption needs defending, because automated program
repair has improved substantially over the same period. Recent research
systems {[}7{]} report plausible patch rates above 80\% on C/C++
vulnerability benchmarks, and commercial tooling {[}8{]} advertises
developer acceptance in the 60 to 70 percent range. If repair costs are
falling as fast as discovery costs, there is no coverage problem to
discuss.

It helps to be precise about what those figures measure. A plausible
patch is one that satisfies the available oracles, typically the
original proof of concept together with the existing test suite. The
benchmarks consist of bugs with reproducible triggers and known-good
tests, which is to say bugs that have already been localised and
characterised. That is the population for which repair was least
expensive to begin with.

The aggregate data runs the other way, and its scope should be fixed
before the numbers are read rather than after. What follows is
deployment coverage, not release coverage: the agent is an operator
rather than a vendor, the population is published vulnerabilities rather
than internal findings, and the cost structure of applying a patch is
not the cost structure of authoring one. The figures therefore establish
a mechanism operating under conditions favourable to remediation, not a
measurement of the quantity this article is about. With that entered:
Verizon's 2026 DBIR, drawing on
more than a billion vulnerability detection records, reports {[}9{]}
that 26\% of vulnerabilities in the KEV catalogue were fully remediated
in 2025, against 38\% the year before, with median time to full
resolution rising from 32 days to 43. The share left entirely untouched
rose from 12\% to 16\% over the same period. These are actively
exploited flaws, so the comparison is not confounded by disagreement
about what counts as severe. The distinction is taken up again in the
conclusion.

Nor is the decline explained by reduced effort. Qualys, a research
partner on the report, observes {[}10{]} that median
detection-to-closure held steady at nine days while KEV-linked workload
grew by 78\%, from 295.8 million instances to 527.3 million. The flat
median is easy to misread. It describes the items that were closed, not
the proportion that were, and that proportion fell. Teams handled each
finding they reached about as quickly as before, and reached a smaller
share of them. The shortfall is also largely independent of maturity:
between 60\% and 70\% of KEV vulnerabilities remain open at day seven
whatever the organisation.

Two conclusions follow, and they rest on different kinds of support. The weaker is that repair improvements have
not yet offset discovery improvements; that is an empirical claim, and
the figures above are evidence for it. The stronger is that they cannot,
so long as patch generation is not the binding step. Blast radius
analysis, regression risk, backporting across
supported branches, release scheduling and customer upgrade cycles are
the expensive parts of remediation, and none of them is a
code-generation problem. Where patch generation is the binding step ---
an architectural change, or one that crosses component boundaries --- it
is binding because the change is hard to author, which is the case
automated repair is furthest from addressing. The stronger conclusion is
argued rather than measured: it rests on an enumeration of where
remediation cost sits, and I am not aware of a published cost breakdown
that would settle the distribution either way. It should be read as a
claim about which step binds, offered for testing, and it fails if
patch authoring turns out to dominate remediation cost in practice.

The same diagnosis indicates where repair research would pay.
Blast-radius estimation, regression risk prediction and automated
backporting are tractable problems, and they attract a fraction of the
attention that patch synthesis does.

One further effect runs against the optimistic reading. If automated
repair clears the well-characterised findings first, the remaining queue
grows more concentrated in the architectural and cross-cutting cases.
The next section argues that these are also the ones correlated with
attacker value. Better repair tooling sharpens the selection effect
rather than offsetting it.

\hypertarget{why-the-unremediated-residue-is-not-a-random-sample}{%
\section{Why the unremediated residue is not a random
sample}\label{why-the-unremediated-residue-is-not-a-random-sample}}

The natural way to model incomplete coverage is as a discount: find a
hundred, fix sixty, capture sixty percent of the available benefit. That
model understates the problem, for three reasons.

The first concerns selection. Remediation queues are ordered by cost
against estimated severity, and cheap fixes clear first. A fix is
expensive when it is architectural, when it crosses component
boundaries, or when it touches something the rest of the system depends
on. Those are approximately the same properties that make the underlying
flaw valuable to an attacker. An outdated dependency carrying a
published CVE in a code path nothing reaches is cheap to fix and worth
nothing to anyone. A trust boundary drawn in the wrong place between two
services is expensive to fix and worth a great deal, because everything
downstream of it inherits the error. The correlation is not clean, and
the clearest counterexample runs the other way: a memory-safety bug in a
parser exposed to untrusted input is usually cheap to fix and commands
high prices, because reachability there does the work that blast radius
does elsewhere. So the claim is the weaker one. Triage sorts on cost,
the expensive end is where the architectural cases sit, and the
surviving residue is biased toward them rather than drawn uniformly from
the population.

The second concerns the artifact. A tracker containing confirmed,
reproducible, severity-rated, unremediated vulnerabilities in shipped
product is more valuable to an adversary than the source code, because
the source code is the input to discovery and the tracker is its output,
already localised, confirmed and rated. At historical discovery rates
the artifact was small enough to overlook. At scanner volumes it is
plausibly the most sensitive document an engineering organisation holds,
and it typically lives in the same issue tracker as everything else,
under the same access controls. Considered as a position, each deferred
entry is a short option written against one's own product: the downside
is unbounded, the counterparty chooses when to exercise, and the premium
--- the ship date it protected and the capacity it released --- was
spent on the day it was collected. Two features of the position are less
obvious than the asymmetry and matter more. The cost of closing it rises
monotonically, because a flaw fixed after shipping carries the
backporting, scheduling and customer-upgrade costs enumerated in the
previous section, none of which existed while the product was still
unreleased; the price of the fix that was declined therefore goes up
every release, and nobody recomputes it. And the position is renewed
without collecting anything. A rolled option at least earns fresh
premium each time; a deferred finding carried into the next release
earns nothing, because the schedule relief was banked once and the entry
simply persists.

The third concerns status. Documented awareness of an exploitable flaw
in a shipped product is discoverable, and it interacts with disclosure
and liability regimes in ways that unrecorded exposure did not. Cheap
scanning manufactures that documentation at volume. It also makes the
asymmetry explicit: the organisation deciding to ship holds the finding,
and the party exposed by that decision does not know it exists. That is
adverse selection in Anderson's sense. The buyer cannot observe what it is being sold, so nothing in
the price disciplines the decision.

Security certification formalised this long before any of it, and did so
explicitly enough to be worth examining. A Common Criteria evaluation is
the clearest published case of an organisation deciding to ship with
known, unfixed vulnerabilities, and of that decision resting on a stated
assumption about the price of an attacker's time. The evaluation
distinguishes exploitable vulnerabilities from residual ones, where a
residual vulnerability is real, identified, unremediated, and judged
exploitable only by an attacker operating above the certified capability
level. The near-collision of terms is unfortunate, and needs
pinning down before it does damage. \emph{Residual vulnerability} is Common
Criteria's term of art for a finding an evaluator has formally priced
and placed out of reach. \emph{Residue}, in this article, is the whole
set of identified and unremediated findings at a release, priced or not,
and for most products no member of it has ever been priced by anyone.
The Common Criteria category is thus a strict subset, and the smaller
one: it is the part of the residue that has been through an evaluation.
Vendors advertise
the outcome, reasonably enough: a determination that the remaining flaws
require an attacker of beyond-high attack potential is a genuine
achievement. It is also a signed certificate asserting that known,
unfixed vulnerabilities are out of reach.

The mechanism by which that judgement is reached is what makes it
relevant now. Attack potential is scored {[}11{]} on the effort a
successful attack requires, including elapsed time and specialist
expertise, which are two of the inputs that a cheap and capable model
most directly compresses. Every such determination encodes an assumption
about what an attacker's hours cost. Certificates are revisited when
products change or lifecycles expire, and not when attackers become
cheaper.

\hypertarget{what-the-regulation-triggers-on}{%
\section{What the regulation triggers
on}\label{what-the-regulation-triggers-on}}

Cheap discovery changes the question an organisation has to answer. It
is no longer why a vulnerability was not found, since increasingly it
was. It is how the decision was made once it had been. Every framework
in the stack could in principle examine that decision. None of them
does.

The EU Cyber Resilience Act gives the situation a concrete legal shape.
Its reporting obligations take effect on 11 September 2026 {[}12{]}.
They require an early warning within 24 hours and a fuller notification
within 72 hours. They also reach backwards, to products already placed
on the market. The remaining requirements, including patching and
remediation, do not apply until 11 December 2027 {[}13{]}.

The trigger is worth attention. It is not discovery, and it is not
internal knowledge. It is active exploitation. A vulnerability found
internally, rated, deferred and signed off attracts no CRA reporting
duty until someone else finds it and uses it. Until 11 December 2027, no
CRA remediation duty attaches to it either. Once exploitation is
reported, market surveillance authorities and national CSIRTs have
visibility, and
continued inaction becomes untenable --- but through instruments other
than the CRA.

The wider legal stack does not fill the gap, because its reporting
duties are constructed the same way. GDPR, NIS2 and the revised Product
Liability Directive all fire on consequences: a personal data breach, a
disruption to essential services, harm caused by a defective product.
Each waits for someone else to exploit what the manufacturer already
documented.

Management duties are a different matter and should be conceded. NIS2
Article 21 requires vulnerability handling by essential entities without
waiting for an incident, and the CRA's own vulnerability-handling
requirements bite from December 2027 on the same basis. But these are
process obligations. They require that a procedure exist and be
followed, not that any particular finding be fixed before any particular
release. The Product Liability Directive is a third case again, reaching
the decision through litigation rather than supervision: a rated,
deferred, documented finding is unusually good evidence of both defect
and foreseeability. What none of them does is observe the specific
decision at the specific release, which is the quantity this article is
about.

A compliance specialist I put this to raised the strongest objection,
which is worth stating plainly. A rational vendor has an incentive to
remediate internally-discovered flaws precisely in order to avoid
triggering the 24-hour escalation, so the regime should pull remediation
forward rather than defer it. I accept the mechanism and dispute its
reach. The incentive determines the order of the queue. Engineering
capacity against a release date determines its throughput. Cheap
discovery raises the input without raising the throughput. The incentive
also prioritises by probability of external discovery rather than by
severity, which directs effort toward shallow and easily-found bugs and
leaves the architectural ones where they were. The objection has a
stronger form that should be conceded. If cheap discovery eventually
raises the probability of external discovery for every class of flaw,
including the architectural ones, the escalation incentive reaches them
too and the misprioritisation corrects itself. But it corrects by
raising the expected cost of the entire residue at once, against
unchanged capacity, which is the constraint this article is about rather
than an escape from it.

The convergence is not accidental. Exploitation is observable to a third
party; internal knowledge is not. A regulator can establish that a
breach occurred, that services were disrupted, that a product caused
harm. It cannot establish what an engineering organisation knew, and
when, without inspecting the organisation. Regimes therefore trigger on
the events they can see, and fall back on process requirements where
they cannot. The residue sits in the gap between the two. One instrument
does reach internal knowledge, and the next section shows why it can.

\hypertarget{the-directive-that-formalises-deferral}{%
\section{The directive that formalises
deferral}\label{the-directive-that-formalises-deferral}}

The evidence has so far moved between two decisions without
distinguishing them. A vendor decides what to ship with; an organisation
running the result decides what to patch and when. The economics are the
same --- fixed capacity, compounding load, triage under a deadline ---
and federal policy has now given the second decision a formal shape.

On 10 June 2026 CISA issued Binding Operational Directive 26-04
{[}14{]}. It supersedes BOD 19-02 and BOD 22-01. The latter was the
directive that had established the KEV catalogue as a remediation
mandate. CISA describes 26-04 as harmonising and improving both, and as
replacing fixed deadlines for federal civilian agencies with a framework
that prioritises high-risk vulnerabilities for timely action while
deferring action against low-risk ones.

Findings are sorted on asset exposure, KEV status, whether an adversary
can automate every step of exploitation, and technical impact.
Satisfying every criterion starts a three-day clock. Lesser findings
receive longer periods. The lowest tier is a defined term, fix on system
upgrade, meaning that the vulnerability should be remediated the next
time the affected asset receives a scheduled major upgrade or rebuild.
The tiers are not fixed. Taking a system off the internet moves a
finding down and a later KEV listing moves it up, so the lowest tier is
a conditional state rather than a resting place.

I take CISA to be correct here. With capacity fixed and load
compounding, prioritisation is the only lever that still moves, and a
directive insisting that everything be patched promptly would be
ignored. The observation is narrower: a federal agency has now
established in a binding directive that a class of identified
vulnerabilities in government systems carries no date-certain
remediation deadline, only an obligation attaching to an event that may
or may not arrive. The residue has ceased to be an operational shortfall
and become policy.

The starting point of the clock is the most consequential detail in the
directive. The timelines begin either when CISA adds the vulnerability
to the KEV catalogue or, under BOD 23-01, when the agency itself
enumerates it on an asset and updates its Continuous Diagnostics and
Mitigation dashboard, whichever comes first. That is a remediation duty
attaching to documented internal identification rather than to
exploitation by someone else, and it is the only such instrument
examined here.

It is not a counterexample to the convergence argument. It is that
argument's mechanism, running forwards. The duty can attach to internal
identification because CISA first mandated the telemetry that makes
internal identification observable: agencies report vulnerability status
automatically through the dashboard, and those that cannot must file
manually every two weeks.

The general form: a regulator's reach over a decision is bounded by its
visibility of the input to that decision. Where the finding is visible,
the decision taken about it can be regulated directly; where it is not,
the regulator is left with the two instruments catalogued in the
previous section --- a trigger on some observable consequence, or a
requirement that a process exist. The convergence described there
follows from what the visibility constraint permits, not from any
failure of legislative imagination.

The sequence matters as much as the mechanism, and BOD 26-04 gets it in
the right order. The telemetry came first, under BOD 23-01, and the
timeline was attached to it afterwards. A duty that attaches to internal
knowledge without an independent record of that knowledge is
self-reported, and a self-reported duty triggered by knowing is exactly
the tax on looking that Rescorla warned about: the compliant response is
to record less, or to look less, and the regulator cannot distinguish
either from success. Because the dashboard is populated independently of
the agency's remediation decisions, looking less is itself visible.
That is what makes the duty enforceable rather than merely stated.

Nothing equivalent exists between a vendor and any authority, which is
why no vendor-facing regime has tried. The condition such a regime would
have to satisfy is now specifiable: a record of findings reaching a
third party without passing through the vendor's editorial control.
Several structures could carry it --- an auditor's access, a
certification scheme's continuing obligation, an escrowed attestation,
or a procurement term --- and they differ in cost and intrusiveness
rather than in kind. None currently exists. Which of them is workable is
beyond this article, but the requirement they must meet is not, and it
supplies a test the argument can fail: a regime that attaches duties to
internal knowledge without first establishing an independent record
should show falling discovery volume among the entities it covers. If
such a regime were introduced and discovery volume held steady, the
mechanism described here would be wrong.

The directive also requires agencies to review their contracts for
modifications needed to meet it. That is the route by which an
operator-side timetable reaches the vendors supplying those operators,
and federal procurement is the fastest instrument discussed in this
article that touches a vendor's release decision. It needs no new
legislation.

Three of the four sorting criteria also deserve notice. Asset exposure,
exploit automation and KEV membership are all proxies for the likelihood
that someone else finds the flaw. The graduated model is right; what
cheap discovery destabilises is the choice of sorting variables, which
were calibrated when external discovery was rare. The directive's own
sorting table is built on
SSVC decision points, and prioritisation by probability of external
discovery is an existing objection to SSVC and EPSS rather than a new
one. What is new is the size of the error it now produces: it sorts a
deep architectural flaw in an unexposed system to the bottom of the
queue. The criteria are also public. An attacker with cheap discovery
can read the directive and infer where the deep and unremediated flaws
are most likely to be.

\hypertarget{why-this-strengthens-rather-than-weakens-the-case-for-verification}{%
\section{Why this strengthens rather than weakens the case for
verification}\label{why-this-strengthens-rather-than-weakens-the-case-for-verification}}

None of the preceding is an argument against formal methods. It is a
reason to regard them as more important than the scanning half of the
response.

Triage is a decision taken repeatedly, indefinitely, under deadline
pressure, and some fraction of those decisions will go the wrong way. A
finding deferred once rejoins the queue behind newer work, and in
practice is rarely reconsidered. The deferral is effectively permanent
even though nobody decided that it should be. Adopting a specification
that rules out a category of bug is a decision taken once, after which
instances cannot be introduced and no triage decision remains to lose.
That is a structural improvement located exactly where the structure is
weakest, which is why the falling cost of proof work seems to me the
more consequential of the two developments.

The failure mode transfers rather than disappearing, however. A
specification adopted against code that violates it produces a failed
proof, at which point the organisation either fixes everything the proof
caught or weakens the specification until it passes. Weakening a
specification under release pressure is the same pathology in a more
respectable form, and it is harder to detect because it presents as
engineering judgement rather than as risk acceptance. Specification
changes should therefore be treated as security-relevant events, subject
to the same review as code. This has to be established as practice
before the first deadline arrives, because by then the question is no
longer neutral.

Two limits on the claim should be stated rather than left to a reviewer.
A proof closes a category that can be specified, and the residue this
article says survives triage is architectural and trust-boundary error,
which is the class where the specification is the thing that was got
wrong. A misplaced trust boundary will be faithfully proved correct
against a specification that encodes the misplacement. Proof therefore
removes the recurring decisions in the specifiable classes, which is
most of the volume even if not most of the value, and leaves the
architectural residue roughly where it was. The partial answer is that
writing a specification is the point at which trust boundaries have to
be stated explicitly, which is earlier than any other schedule forces
the question. The second limit is that adopting verification is itself a
decision taken under the capacity constraint just diagnosed. Nothing
exempts it. It only has to survive being made once.

\hypertarget{where-the-argument-does-not-reach}{%
\section{Where the argument does not
reach}\label{where-the-argument-does-not-reach}}

Chlipala names legacy systems and untrained teams as the outstanding
challenge and defers them to later work. My only substantive
disagreement concerns magnitude. I do not think this is a transition
cost on the way to a solved state; I think it is the dominant term.

Both halves of the response require infrastructure. Pre-release scanning
requires a pipeline under one's control. Verification requires code
written against a specification. Neither describes the software
currently running most of the world, which is precisely the population
from which the allocational subsidy has been withdrawn. The withdrawal
happens on the attacker's schedule and the transition runs on a
decade-long one. The gap between those two clocks is the exposure, and
no improvement in either technique closes it. What remains is managing
the residue they leave, which begins with measuring it.

\hypertarget{what-follows-for-measurement-and-effort}{%
\section{What follows for measurement and
effort}\label{what-follows-for-measurement-and-effort}}

If coverage at the release decision is the binding constraint, most
security programmes are reporting the wrong quantity. The practitioner
literature has noticed part of this already {[}15{]}: tracking total
vulnerabilities found creates perverse incentives, rewarding detection
volume over risk reduction and encouraging teams to add scanners rather
than fix findings. The observation is correct and does not go far
enough. Neither counting findings nor mean time to remediate reports the
fraction that was not fixed.

The same reasoning applies to where effort goes, not only to what gets
reported. If discovery is no longer the binding step, a further
increment of discovery capability buys less than an equivalent increment
in conversion. This is a claim about margins rather than about worth:
the argument says nothing against the first scanner, and something about
the fifth.

One qualification matters. A discovery improvement that moves a finding
earlier --- into design review, or into the editor --- is a conversion
improvement in other clothes, because cost-to-fix rises with distance
from the decision that introduced the flaw. What has diminishing returns
is detection volume at a fixed point in the pipeline.

Three different things can be done about the residue, and they are easy
to confuse. It can be reduced, by converting more findings and by moving
findings earlier. It can be governed, by controlling who sees the
backlog, who accepts what remains, and when the acceptance expires. And
its management can be demonstrated, through the documentary record a
certification scheme, a supervisory authority or a court would examine.
The convergence argument above explains why the third is the one under
pressure: it is the only one an outside party can observe, which makes
it both the cheapest to invest in and the easiest to substitute for the
first. Documenting a deferral well is not a substitute for not
deferring.

Five things follow, ordered by who has to act on them.

Report coverage per release rather than discovery volume: findings
raised against findings remediated, with the deferred remainder visible
to whoever signs the release. Attaching it to the release decision means
the number has to move something: a threshold below which the release
does not proceed unchanged, or a named person who accepts the remainder
in writing, with a date on which it returns. Report it as
Equation~\ref{eq:coverage} requires, with the scan scope and
configuration declared alongside the ratio, since a figure without a
declared scope can be improved by looking less. It is what makes the
deferral countable, and what makes a pattern of them visible before the
pattern becomes the design. It is also the figure a buyer can ask for,
which is the point the conclusion returns to.

Treat the deferred backlog as a sensitive asset, with access controls,
monitoring and retention policy appropriate to a document that maps
exploitable flaws in shipped product.

Re-examine severity and attack-potential judgements made when an
attacker's time and expertise were expensive. This is an assurance
problem as much as an engineering one. Every such judgement encodes an
assumed price for an attacker's hours, and that price should be recorded
rather than left implicit, so that it can be checked when it stops
holding. Certification schemes revisit determinations when a product
changes or a lifecycle expires, and nothing triggers a revisit when the
assumption underneath the determination expires instead. It must
therefore be scheduled deliberately.

Decide a disclosure posture, and decide it deliberately. Every regime
that reaches vendors fires on exploitation by someone else, so the
distance between what an organisation knows and what it must report is
now set entirely by the organisation. The question of whether a duty
should instead attach to documented internal knowledge is a real one,
and the objection to it is Rescorla's: a duty triggered by knowing is a
tax on looking, and the cheapest way to comply is to look less. Any
regime that closes this gap has to solve that problem first. BOD 26-04
shows what a solution looks like: its duty attaches to a record the
regulator already receives through mandated telemetry, so compliance is
not self-reported and looking less is itself visible. Whether that
generalises from federal operators to vendors is the open question.
Until it does, the gap is a choice each vendor is making silently.

Begin verification on new code, and defend the specification once
adopted, for the reasons given earlier. It is listed last because it is
the slowest to arrive, not because it matters least.

\hypertarget{conclusion}{%
\section{Conclusion}\label{conclusion}}

Discovery costs have fallen and will fall further; so have patch
generation costs, and neither trend has translated into higher release
coverage. As argued in Section 3, patch generation was never the binding
step. What determines exposure is
the fraction of identified vulnerabilities remediated before shipping,
and neither cost curve directly changes that fraction.

Two populations have to be kept apart here, because they have changed in
different ways. Flaws that had been found and deferred are simply more
numerous than they were, which is a change in volume. Flaws that nobody
had looked for are the larger change. They carried no epistemic status
at all --- no ticket, no rating, no signatory, nothing a court or a
buyer could examine --- and cheap discovery gives them one. That is a
change in kind, and it is what this article's title names. The artifact
and status arguments of Section 4 are about that migration; neither is
new for a flaw that already sat in someone's tracker. The change in
volume is not thereby uninteresting: a backlog small enough to work
through is work outstanding, and one large enough that a federal
directive formalises indefinite deferral has become a structural feature
of how software ships.

Release coverage, as defined in Section 2, is not measured, and no
vendor-side figure exists.
Establishing one is part of what this article is asking for. The nearest
proxy available is deployment coverage --- a different agent and a
different cost structure, so what transfers is the mechanism rather than
the number --- in the easiest case imaginable: actively exploited flaws,
listed in a public catalogue, in systems already deployed. There, full
remediation stands at 26\%, against 38\% the year before, with the share
left entirely untouched rising from 12\% to 16\%. If remediation falls
that far short under those conditions, findings known only to the
organisation
that recorded them face no external pressure at all, and have no reason
to do better.

Neither pre-release scanning nor verification reaches the software
already running: scanning needs a pipeline, proof needs a specification.
For everything else, what remains is the decision taken at each release
about what to ship with, and the first thing to do about that decision
is to count it.

Counting it is also the precondition for anything else. The residue sits
with the vendor and the exposure sits with the user, which is the
externality Anderson described: the party taking the decision is not the
party that pays for it, and nothing in the price of the software
reflects the difference. Externalities are internalised by liability, by
insurance, or by contract, and all three need a number that can be
stated at the point of sale. Liability waits for harm and then argues
about it. Insurance would have to price a quantity nobody currently
reports.

Procurement needs neither precondition. A buyer can ask for release
coverage today,
and BOD 26-04 has already instructed federal agencies to reopen their
contracts. That is the fastest route available, and nobody has to wait
for a regulator to act.

The older literature asked what we were buying with the hours we spent
looking. The answer now depends less on the looking than on what we do
with what we find. Cheap discovery has not removed the economics of
vulnerability; it has moved them, from the scarcity of finding flaws to
the scarcity of acting on the ones already found. For software that
reaches a release decision, the binding boundary is no longer discovery
but that decision, and release coverage is what makes the boundary
visible.

\hypertarget{references}{%
\section*{References}\label{references}}
\addcontentsline{toc}{section}{References}

\begin{enumerate}
\def\labelenumi{\arabic{enumi}.}
\tightlist
\item
  Anthropic Red Team. Claude Mythos Preview.
  \url{https://red.anthropic.com/2026/mythos-preview/} Accessed 5 August
  2026.
\item
  UK AI Security Institute. Our evaluation of Claude Mythos Preview's
  cyber capabilities.
  \url{https://www.aisi.gov.uk/blog/our-evaluation-of-claude-mythos-previews-cyber-capabilities}
  Accessed 5 August 2026.
\item
  OpenAI. Hugging Face model evaluation security incident.
  \url{https://openai.com/index/hugging-face-model-evaluation-security-incident/}
  Accessed 5 August 2026.
\item
  Anderson, R. Why Information Security is Hard --- An Economic
  Perspective. 2001. \url{https://www.cl.cam.ac.uk/~rja14/econsec.html}
  Accessed 5 August 2026.
\item
  Rescorla, E. Is Finding Security Holes a Good Idea? Workshop on the
  Economics of Information Security, 2004.
  \url{http://www.dtc.umn.edu/weis2004/rescorla.pdf} Accessed 5 August
  2026.
\item
  Chlipala, A. The End of Security Through Obscurity.
  \url{https://stng.substack.com/p/the-end-of-security-through-obscurity}
  Accessed 5 August 2026.
\item
  Wang, H., Basque, Z. L., Hu, J., et al. Root-Cause-Driven Automated
  Vulnerability Repair. arXiv:2605.04251, 2026.
  \url{https://arxiv.org/html/2605.04251v1} Accessed 5 August 2026.
\item
  Pixee. Best automated remediation tools, 2026.
  \url{https://www.pixee.ai/blog/best-automated-remediation-tools-2026}
  Accessed 5 August 2026.
\item
  Veracode. 2026 Verizon DBIR: application security.
  \url{https://www.veracode.com/blog/2026-verizon-dbir-application-security/}
  Accessed 5 August 2026.
\item
  Qualys. Inside the 2026 Verizon DBIR: what one billion records
  revealed about vulnerability remediation.
  \url{https://blog.qualys.com/vulnerabilities-threat-research/2026/05/19/inside-the-2026-verizon-dbir-what-one-billion-records-revealed-about-vulnerability-remediation}
  Accessed 5 August 2026.
\item
  ENISA. Application of attack potential to smartcards.
  \url{https://certification.enisa.europa.eu/publications/application-attack-potential-smartcards_en}
  Accessed 5 August 2026.
\item
  European Commission. Cyber Resilience Act reporting obligations.
  \url{https://digital-strategy.ec.europa.eu/en/policies/cra-reporting}
  Accessed 5 August 2026.
\item
  Crowell \& Moring. EU Cyber Resilience Act countdown: 11 September
  2026 incident and vulnerability reporting deadline.
  \url{https://www.crowell.com/en/insights/client-alerts/eu-cyber-resilience-act-countdown-11-september-2026-incidentvulnerability-reporting-deadline-is-less-than-100-days-away}
  Accessed 5 August 2026.
\item
  CISA. Binding Operational Directive 26-04: prioritizing security
  updates based on risk.
  \url{https://www.cisa.gov/news-events/directives/bod-26-04-prioritizing-security-updates-based-risk}
  Accessed 5 August 2026.
\item
  AppSec Santa. AppSec metrics guide.
  \url{https://appsecsanta.com/application-security/appsec-metrics-guide}
  Accessed 5 August 2026.
\end{enumerate}

\end{document}